\documentstyle[preprint,eqsecnum,prd,aps]{revtex}
\begin{document}
\draft

\title{  QED symmetries in real-time thermal field theory. }
\author{ Juan Carlos D'Olivo$^a$\footnote{e-mail: dolivo@aurora.nuclecu.unam.mx},
Manuel Torres$^b$\footnote{e-mail: manuel@teorica1.ifisicacu.unam.mx}  and
Eduardo T\'ututi$^b$\footnote{e-mail: tututi@teorica0.ifisicacu.unam.mx}}
\address{ $^a$ Instituto de Ciencias Nucleares, 
Universidad Nacional   Aut\'onoma    de M\'exico \\
Apdo. Postal 70-534, 04510  M\'exico, D.F., M\'exico.  \\
$^b$ Instituto de F\'isica,   Universidad Nacional  Aut\'onoma de  M\'exico \\  
Apdo. Postal 20-364,  01000  M\'exico, D.F., M\'exico. }

\maketitle
\begin{abstract}
We study  the  discrete and gauge symmetries  of  Quantum Electrodynamics at  finite temperature  within the  the real-time formalism.  
The gauge invariance of the complete generating functional  leads to the finite temperature Ward  identities.  These Ward identities  relate the eight vertex functions to the elements  of the self-energy matrix. 
Combining the relations  obtained from the $Z_2$  and the gauge symmetries 
of the theory we find that  only one out of  eight longitudinal vertex functions is independent.
As a consequence  of the  Ward identities it is shown that 
some elements of the vertex function are  
singular when the photon momentum goes to zero.
\end{abstract}

\narrowtext

\section{Introduction}\label{introduction}

Quantum field theory  at finite temperature  (QFFT)   has been 
a  subject  of  great current  interest for  a variety of physical contexts 
\cite{gen}. 
Two main formulations of QFFT can be distinguished.  In the 
imaginary-time formalism (ITF) \cite{matsu}  the   continuous energy variables $k_0$  are replaced   by the  discrete
Matsubara  frequencies $\omega_n$, and  loop integrals over  $k_0$ are  transformed into discrete sums.  
To  obtain a  graph  with real external energies  requires  a non-trivial  analytic continuation.   On the other hand,  
the real-time  formalism   (RTF)    allows the computation of dynamical quantities directly as functions of continuous real frequencies.    
In addition,  it is possible to formulate the theory in a  covariant way \cite{gen}.
The price to  paid is that 
the   Feynman  rules are now  more complicated that those  at zero temperature.   The number of degrees of freedom  is doubled and hence the  propagator acquires a matrix structure.    The RTF has been discussed using a canonical operator 
formulation \cite{ume} and also  using the  path integral method \cite{niemi}.  Here,  we shall mainly  consider the diagrammatic method based in  the path integral representation of the generating functional. 

One would expect physical quantities to be independent of the formalism used to calculate them.  This, however has been an issue  that has generated a longstanding controversy.   For  example,     the amputated  $n$-point  functions  at  finite  temperature   appear  to lead to different results when calculated in the two formalism's.
 Nevertheless,  it has been shown that   within the real-time formalism one can consider a sum of graphs, motivated by causality arguments, which at least for the two- and three-point functions agree with the corresponding analytically continued imaginary-time results
\cite{evans,kobes3,kobes}. Thus,  whereas  single graphs may differ in the two formalism's, there exist   quantities that agree when calculated by either method.

In this work we are interested in studying the discrete and gauge symmetries of  QED at finite temperature within the real-time formalism. 
It is commonly believed that gauge invariance is not affected
by the temperature \cite{ume}. There exist, of course,  some previous 
studies on the subject  \cite{koku,carri2}; in particular, in the case of  gauge theories at high temperature it has been  shown that the  hard thermal loops  and the tree amplitudes obey the same Ward identities \cite{braten}. That  Ward  identities may be carried over to finite temperature without any modification, except for the fact that the matrix elements  depend on the temperature,  is probably obvious in the  ITF.  However  in the RTF, where  there are eight different   vertex,  the structure of the Ward identities is far from obvious.  To our  knowledge  an explicit  derivation of the Ward identities in the RTF  has not appeared in the  literature.  The purpose of this article is to 
provide a systematic derivation based on the path integral method.   In particular, we consider the Ward identities that  relate the 2- and 3- point Green functions.

Section  \ref{qeds} is devoted to the  study of the effective two component thermal formulation of  QED within the path integral formalism's.  Along the paper  we stress the fact  that the thermal information in the  action is  contained in the convergence factor $\epsilon$ that is inserted through boundary conditions.
We also  discuss the discrete symmetries of the effective $QED$ action.
In section  \ref{wardsec} the subject of the  gauge invariance of the theory  is examined.  Ward identities are obtained by demanding
 the complete generating functional  be gauge invariant.  
When  applied to the photon propagator, these identities  imply that any component of the photon  polarization tensor is transverse to any order of perturbation theory.  Furthermore, the Ward identities  impose several relations among  the  vertex functions and the  elements  of the self-energy matrix. We exploit these relations  to prove that 
only one  out of   eight components of the longitudinal part of the vertex 
functions is independent. We also show that the Ward identities at finite temperature demand that some  of the vertex functions have to be singular  when the photon momentum goes to zero.  Finally,  the  general results are illustrated 
by  a one-loop calculation.
 
\section{Thermal QED}\label{qeds}

Let us  consider a gas of electrons and photons  at 
finite temperature. The system is described by the  
 $QED$ Lagrangian 
\begin{equation}\label{lqed}
{\cal L} = -{1 \over 4}{F}^{\mu\nu}F_{\mu\nu}+i\bar{\psi}
\gamma^{\mu}(\partial_{\mu}+ieA_{\mu})\psi - m\bar{\psi}\psi
 - { \xi \over  2}(\partial \cdot A )^2
\, , 
\end{equation}
which includes the gauge fixing term.
The path integral representation of the generating functional can be written as
 
\begin{equation}\label{zq1}
Z\,\left[J_\mu, \eta,\bar{\eta}\right] = \int[D A_\mu D \psi D \bar{\psi}]
\, \exp \left \{ i {\int_{C}} d\tau d \vec{x} \left({\cal L}  +\bar{\psi} \eta +
\bar{\eta} \psi + J_\nu A^\nu \right) \right\} \, , 
\end{equation}
where  the  integration over $\tau$ is along a 
monotonically descendent  contour $C$ in the 
complex time-plane. The contour has end points $\tau_i$ and 
$ \tau_i - i \beta $, with $\beta $ denoting  the inverse of the temperature.
The fields in (\ref{zq1}) satisfy the boundary conditions
 $A_\nu(\tau_i) = A_\nu(\tau_i - i \beta) \, $, and 
 $\psi(\tau_i) = - e^\mu \psi(\tau_i - i \beta)$,  where $\mu$ is the chemical potential.
Except for  the gauge-fixing  and the source terms  the  action in (\ref {zq1}) is gauge invariant.
Indeed,  the first three terms in  ${\cal{L}}$ are invariant  under the  gauge transformation 

\begin{equation}\label{tn1}
A_\mu \rightarrow A_\mu  + \partial_\mu \Lambda
\, , \qquad \psi \rightarrow \exp\left\{- i e \Lambda \right\} \, \psi \, ,
\end{equation}
with the time derivative taken along $C$.

Within the  RTF  the curve $C$ is chosen  in such a way that  includes the  real time axis.
 A particular  contour  that has been extensively used is one  of the form 
$C = C_1 \oplus C_2 \oplus C_3 \oplus C_4 $ . The segment $C_1$  
starts at $\tau_i  = - T$ and runs along the real time axis up  to $ +T$.   $C_2$  continues along the imaginary axis from $+ T$ to $ + T - i \sigma \beta$,  $C_3$  goes from $+T  - i \sigma \beta$ 
to     $-T  - i \sigma \beta$, and  $C_4$  runs from    $-T  - i \sigma \beta$ to 
$-T - i \beta$. At some stage, the   limit  $T \to \infty  $  is taken to produce the standard RTF. The contour parameter $ \sigma $ can take any value within the interval $(0,1)$. Here we shall adopt the symmetrical value $\sigma = 1/2$.
Of course, physical results should not depend on $\sigma$. 
 
Following the method of Niemi and Semenoff \cite{niemi}  $Z$ can be recast as an integral over fields evaluated  at real time,  but with the 
number of degrees of freedom effectively doubled. 
The fields lying on  $C_1$   (type-1 field)
are referred as the physical field, while the (type-2) fields associated with the
section $C_2$ of the contour are named  as (thermal) ghost fields. 
In terms of these fields 
the generating functional   takes the form \cite{kose}:

\begin{equation}\label{zq2}
Z\left[J^\mu_a, \eta_a,\bar{\eta}_a\right] = \int[D A^\mu_a D \psi_a D \bar{\psi}_a]
 \exp \left \{i S_{eff}  \right\}  \,  ,
\end{equation}
where the effective action 

\begin{equation}\label{sef}
S_{eff} =   \int  \left( {\cal L}[1,2] + \bar{\psi}_a \eta_a + \bar{\eta}_a \psi_a 
+ J_{\nu a} A^\nu_a \right) \, d x   \, , 
\end{equation}
has been written in terms of the  finite-temperature Lagrangian 
$ {\cal L}{[1,2]}  = {\cal L}_\psi + {\cal L}_A  +   {\cal L}_{int} $, 
with
\begin{eqnarray}\label{lef}
{\cal L}_\psi  &=&   \bar{\psi}_a \left[ S^{(0)} \right]^{-1}_{ab}  \psi_b  
\,  , \nonumber \\
{\cal L}_A &=&  {1 \over 2} A^\mu_a \left[ D_{\mu\nu}^{(0)} \right]^{-1}_{ab}
 A^\nu_b \,  , \nonumber \\
 {\cal L}_{int} &=&  - \left[ e \bar{\psi}_1 \gamma_\mu \psi_1 A^\mu_1  - 
 e \bar{\psi}_2 \gamma_\mu \psi_2 A^\mu_2 \right]  \, .
\end{eqnarray}
In these equations a  summation over the repeated index $a =1,2$ is understood, 
while   $ S^{(0)} _{ab}$ and 
$\left[D_{\mu\nu}^{(0)} \right]_{ab}$ denote the matrix elements of the
bare thermal   propagators.  At the tree level,  there are only  two  kinds  of  
vertices,  the type-1  and the  type-2 vertices, that differ by a sign: 
$\Gamma^\mu_{111} =  - \Gamma^\mu_{222} = \gamma^\mu$. 
The physical Green functions can be expressed  as  a sum 
of  Feynman diagrams, with  the type-1 fields appearing on the external legs; the 
type-2 field  only occur in the internal lines of the diagrams.  

In momentum space the inverse of the  fermion  propagator can be  worked out as  

\begin{equation}\label{invf}
\left[ {\bf  S}^{(0)}  (p) \right]^{-1}  = \left( {p} {\mkern -9.0mu /} - m  \right) \sigma_3 + i \epsilon 
\left[ {\bf  V}^{-1}(p_0)  \right]^2 \, , 
\end{equation}
where $\sigma_3$ is the diagonal Pauli matrix and 

\begin{equation}\label{mv}
 {\bf V } = 
\pmatrix{
\cos  \varphi_p  & - \epsilon(p_0)  e^{\beta \mu/2} \sin  \varphi_p\cr
\epsilon(p_0)  e^{-\beta \mu/2}\sin \varphi_p  &  \cos  \varphi_p \cr
}  ,  \qquad 
\cos \varphi_p =
 {\theta(p_0) e ^{x/4}  + \theta(-p_0) e ^{- x/4}  \over \sqrt{e^{x/2} + e^{-x/2}} }
 ,  
\end{equation}
with $x = \beta(p_0 - \mu)$.  For the photon the result is 

\begin{equation}\label{invg}
\left[ {\bf  D}^{(0) }_{\mu\nu}  (k) \right]^{-1} = \left[ - k^2 g_{\mu\nu} + ( 1 - \xi) k_\mu k_\nu  \right]
\sigma_3 - i \epsilon g_{\mu\nu}  \left[ {\bf  U}^{-1}(k_0) \right]^2 \, ,
\end{equation}
with 
\begin{equation}\label{mu}
{\bf U }= 
\pmatrix{
\cosh \theta_k  & \sinh \theta_k\cr
\sinh \theta_k  &  \cosh \theta_k \cr
} \,  , \qquad  \cosh^2 \theta_k  =  {1 \over 1 - \exp^{- \beta |k_0|}}
\, . 
\end{equation}

We have  kept  $\epsilon$ finite  in order to define the theory properly \cite{gen}. At zero temperature  $\epsilon$  is included as  a convergence factor, so the path  integral in Minkowski space is well defined. At finite temperature the procedure becomes  essential,  not only for convergence, but also  to keep trace of the temperature dependence, which as seen from the previous expression  appears only in the $\epsilon$-terms.
Because  of these contributions, the action is non-local in the time variable. In fact, 
the substitution of (\ref{invf})  in (\ref{lef}) leads to the following expression for  the fermion part of the action  in   configuration space  

\begin{equation}\label{sexpf}
\int {\cal L}_\psi= \int dx_0 dy_0 d \vec x  \, 
  \bar{\psi}_a(x_0,\vec x) 
\left [ \delta(x_0 - y_0) \left( i {\partial}{\mkern -9.0mu /} - m  \right) \sigma_3
 + i \epsilon {\bf  M}(x_0 - y_0)  \right]_{ab} \psi_b (y_0,\vec x)
\, , \end{equation}
where 
\begin{equation}\label{mm}
{\bf M}_{ab}(x_0 - y_0)  = \int {dk_0 \over 2 \pi }  \left[ {\bf  V}^{-1}(k_0)  \right]^2_{ab}
e^{ - i k_0 (x_0 - y_0)}  \, .
\end{equation}
Similarly,  for the gauge-field  we obtain

\begin{eqnarray}\label{sexpg}
\int {\cal L}_A &=& 
\int dx_0 dy_0 d \vec x  \,  A^\mu_a(x_0,\vec x)  \times \nonumber \\
&& {1\over 2} \left[ \delta(x_0 - y_0) \left\{ g_{\mu\nu} 
{\,\lower 1.0pt\vbox{\hrule \hbox{\vrule height 0.3 cm \hskip 0.28 cm \vrule height 0.3 cm}\hrule}\,}
   + (\xi - 1) \partial_\mu  \partial_\nu  \right\} \sigma_3   +
 i \epsilon g_{\mu\nu} {\bf  N}(x_0 - y_0)
\right]_{ab} A^\nu_b (y_0,\vec x) \,,  \nonumber \\
\end{eqnarray}
with 
\begin{equation}\label{mn}
{\bf N}_{ab}(x_0 - y_0)  = \int {dk_0 \over 2 \pi }  \left[ {\bf  U}^{-1}(k_0)  \right]^2_{ab}
e^{ - i k_0 (x_0 - y_0)}  \, . 
\end{equation}
In the  zero temperature limit  ${\bf N} = {\bf M}  = {\bf I}  \delta(x_0 - y_0)$,  
where $ {\bf I}  $ is the identity matrix.  Then,  the  physical and ghost fields uncouple and the  time-locality  of the action is restored.  

In reference \cite{niemi} it  was pointed out   that the scalar theory at finite temperature has a $Z_2$ symmetry under the interchange of  type-1 and 
type-2 fields. Here  we exploit  the $Z_2$ symmetry for thermal  $QED$ 
to reduce the number 
of independent  $2$- and $3$-point vertex functions.    
 As usual we  introduce the generating functional of one particle irreducible graphs

\begin{equation}\label{gam}
\Gamma \left[ \bar{\Psi}_a , \Psi_a,  A^\mu_a \right]  = 
W \left[\bar{\eta}_a,  \eta_a  , J^\mu_a \right]  - \int \left(
 \bar{\eta}_a  \Psi^a +  \bar{ \Psi}_a  \eta^a +
J^\mu_a  A_\mu^a  \right) \, ,
\end{equation}
where $W \left[\bar{\eta}_a,  \eta_a  ,J_a^\mu \right]  = - i  \ln 
Z \left[\bar{\eta}_a,  \eta_a  ,J^\mu_a \right] $ is the connected generating functional and
 
\begin{eqnarray}\label{rel1}
{ \delta W \over \delta J^a_\mu  } &=&  A^a_\mu 
 \, , \qquad { \delta  \Gamma  \over \delta  A^a_\mu }  = - J^a_\mu 
\, , \nonumber \\ \nonumber \\ 
{ \delta W  \over \delta \bar{\eta}^a  } &=&  \psi^a  \, , \qquad
 { \delta  \Gamma  \over \delta \psi^a  }  = - \bar{\eta}^a 
\, , \nonumber \\ \nonumber \\ 
{ \delta  W \over \delta  \eta^a } &=&  \bar{\psi}^a 
 \, , \qquad { \delta \Gamma  \over \delta  \bar{\psi}^a  }  = -  \eta^a 
\, . \end{eqnarray}
In Eqs. (\ref{gam}) and (\ref{rel1}), ${\Psi}_a$ , $ \Psi_a$, and $A^\mu_a$ refer to the classical fields,  but  for  simplicity we have denoted them with the same notation  than the  one used 
for the fields  appearing in the path integral.
  It is convenient to write the fields as two component spinors 
\begin{equation}\label{2cf}
{\bf A}^\mu = \left (\matrix{ A^\mu_1\cr A^\mu_2\cr }\right ) \, , \qquad  
{\bf \Psi} = \left (\matrix{ \psi_1\cr  \psi_2\cr }\right ) \, , \qquad 
 \bar{{\bf \Psi}} = \left (\matrix{ \bar{\psi}_1 & \bar{\psi}_2   \cr
}\right )   \, .
\end{equation}
In terms of them the $Z_2$ symmetry
for $\Gamma$   is realized  as follows:

\begin{equation}\label{simg}
\Gamma[ \bar{{\bf \Psi}} , {\bf \Psi}, {\bf A^\mu} ] = 
- \Gamma^*[ \bar{{\bf \Psi}}^* \tilde{\sigma}  ,  \tilde{\sigma} {\bf \Psi}^*,
\sigma_1 {\bf A^\mu}]  \, ,
\end{equation}
where $\sigma_1$ is  the Pauli matrix and 

\begin{equation}\label{con2}
    \tilde{\sigma}= 
\pmatrix{ 0 & i e^{ \beta \mu/2} \cr -  i e^{-  \beta \mu/2 } & 0 \cr } \, . 
\end{equation}

As  is well known, the fermion  $2$-point  function 
$\delta^2 \Gamma/\delta \bar{\psi} \delta\psi$  can be related to the inverse of the exact propagator. The explicit relation in  momentum space is 

\begin{equation}\label{2pv}
\left( 2 \pi \right)^4 \delta(p - p^\prime) i S^{-1}_{ab}(p) =
\int dx dy  e^{i \, (p^\prime x - p y)}  {\delta^2 \Gamma
\over \delta \bar{\psi}_a (x)  \delta\psi_b(y) } \, . 
\end{equation}
Taking into account that
${\bf  S} ^{-1}   (p) =  \left[ {\bf  S}^{(0)}  (p) \right]^{-1}  -  {\bf \Sigma}  (p)$,
the $Z_2$ symmetry   (\ref{simg}) imply that  the elements  of the fermion self-energy matrix ${\bf \Sigma}$ are  related as follows:
$\Sigma_{11}(p) =- \Sigma_{22}^*(p)$,  $ \Sigma_{12}(p) = e ^{\beta\mu} \Sigma^*_{21}(p)$.  An additional relation among the elements of 
of ${\bf \Sigma}$ is obtained from the circling relation 
 \cite{kobes3}:

\begin{equation}\label{lgt}
\Sigma_{11}(p)  +   e^{- \beta p_0/2}  \, \Sigma_{12}(p)  +
 e^{ \beta p_0/2} \,  \Sigma_{21}(p)  + \Sigma_{22}(p)
 \, = \, 0  \, .\end{equation}
The previous relations taken together  imply that   only one component of the matrix self-energy is independent.  Therefore, for example, knowing $\Sigma_{11}$ is enough to determine the other components of  ${\bf \Sigma}$. In particular,  for the retarded self-energy,  $\Sigma_R (p) = \Sigma_{11}(p) + e^{-\beta p_0 /2} \Sigma_{12}(p)$,  we have :

\begin{eqnarray}\label{sigret}
{\rm Re} \, \Sigma_R (p) \, &=& \, {\rm Re} \Sigma_{11}(p) \, \nonumber \\
{\rm Im} \, \Sigma_R(p) \, &=&   \epsilon(p_0) \sec (2 \varphi_p ) \, {\rm Im} \Sigma_{11} (p)
 \, = \, -i e^{-\beta \mu/2} \csc (2 \varphi_p)  \,  \Sigma_{12}(p) \,  .
\end{eqnarray}
In practice it is simpler to calculate  ${\rm Im} \Sigma_R$ in terms
of the  $\Sigma_{12}$, which  is a purely imaginary quantity . For this reason, 
and also  for future comparison with the result for the 3-point
function (see Eq. (\ref{3pvret})),  
we have written the second equality for the imaginary part of $\Sigma_R$.
 
We now turn the attention on the proper vertex function.
Within the RTF there are  eight  connected  three point functions  $\Gamma^\mu_{abc}(q,p)$, which  are defined by 

\begin{equation}\label{3pv}
i e \left( 2 \pi \right)^4 \delta(p^\prime - p - q )
\Gamma^\mu_{abc}(q,p) =
\int dx dy  dz e^{i \, (p^\prime z - p y - q x)}  {\delta^3 \Gamma
\over \delta \bar{\psi}_c (z)  \delta\psi_b(y)  \delta A^\mu_a(x)} \, .
\end{equation}
The first index corresponds to the photon with momentum $q$, while the second and the third indices  refer to the incoming  fermion with momentum $p$  and the outgoing fermion with momentum $p + q$, respectively.  Not all  of these complex functions  are independent.  The $Z_2$ symmetry of the theory  (\ref{simg}) leads to the following  relations: 

\begin{eqnarray}\label{rel2}
\Gamma^\mu_{222}(q,p) &=& -  \left[\Gamma^\mu_{111}(q,p)\right]^*
 \, , \nonumber \\
\Gamma^\mu_{221}(q,p) &=& e^{\beta\mu} \left[\Gamma^\mu_{112}(q,p) 
\right]^* \, , \nonumber \\
\Gamma^\mu_{212}(q,p) &=& e^{-\beta\mu} \left[ \Gamma^\mu_{121}(q,p)
\right]^*  \, , \nonumber \\
\Gamma^\mu_{122}(q,p) &=& - \left[\Gamma^\mu_{211}(q,p)\right]^*  \, .
\end{eqnarray}
The circling equation of Ref. \cite{kobes3} adapted  to  the vertex function of 
$QED$  introduces a further relation among  the  various $\Gamma^\mu_{abc}$, namely

\begin{eqnarray}\label{rel3}
 \Gamma^\mu_{111}  &+&  e^{- \beta q_0/2}  \Gamma^\mu_{211} + 
 e^{- \beta p_0/2}  \Gamma^\mu_{121}  + e^{- \beta r_0/2}  \Gamma^\mu_{221}
\nonumber \\
&+& e^{ \beta r_0/2}   \Gamma^\mu_{112}  +  e^{ \beta p_0/2}  \Gamma^\mu_{212} +  e^{ \beta q_0/2}  \Gamma^\mu_{122}  +  \Gamma^\mu_{222}
\, = \, 0  \, , 
\end{eqnarray}
where $r_0 = p_0 + q_0$. Equations  (\ref{rel2}) and (\ref{rel3}) represent five complex constrictions and   consequently, one is left  with only three independent   3-point functions.  It is convenient  to choose them in  the following way   \cite{kobes3}:
  
\begin{eqnarray}\label{3pvr}
\Gamma^\mu_{R1}(q,p)  &=& \Gamma^\mu_{111}  +  e^{- \beta p_0/2}  \Gamma^\mu_{121} + 
 e^{ \beta r_0/2}  \Gamma^\mu_{112}  + e^{\beta q_0/2}  \Gamma^\mu_{122}
\, , \nonumber \\ 
\Gamma^\mu_{R2}(q,p)  &=& \Gamma^\mu_{111}  +  e^{- \beta q_0/2}  \Gamma^\mu_{211} + 
 e^{ \beta r_0/2}  \Gamma^\mu_{112}  + e^{\beta p_0/2}  \Gamma^\mu_{212}
\,  , \nonumber \\ 
\Gamma^\mu_{R3}(q,p)  &=& \Gamma^\mu_{111}  +  e^{- \beta q_0/2}  \Gamma^\mu_{211} + 
 e^{- \beta p_0/2}  \Gamma^\mu_{121}  + e^{- \beta r_0/2}  \Gamma^\mu_{221}
\,  . \end{eqnarray}
These are the retarded functions with, respectively,  $t_1$, $t_2$, and $t_3$
as the largest time,  their complex conjugate being the
corresponding advanced functions.
 In the next section we shall see how the Ward identities lead to a further 
reduction in the number of independent 3-point functions. 

\section{Gauge invariance and Ward identities}\label{wardsec}

As stated  in  section  \ref{qeds},  at  finite temperature  the Lagrangian  (without gauge-fixing)  with support  in the  complex contour  $C$   is gauge invariant. However,
the original  theory has been  replaced by an effective one with two components, and  we must ask about the gauge-invariance properties
of this effective theory, as well as  of their consequences. To investigate this point,
let us consider  the following independent gauge transformations on  the  type-1 and type-2 fields:   

\begin{eqnarray}\label{tn2}
A_a^\mu &\rightarrow A_a^\mu  +\partial^\mu \Lambda_a  \, , \nonumber\\
\psi_a &\rightarrow {\rm exp}( - i e \Lambda_a) \psi_a  \, ,
\end{eqnarray}
where $\Lambda_a(x)$ ($a=1,2$) are arbitrary functions of $x$,
that in what follows we assume to be infinitesimal. 
Under such transformation, 
the  finite temperature  effective action  $S_{eff}$ in  (\ref{sef}) is not  invariant; the  origin  is   threefold: 
The same  as in the zero temperature case,   the gauge fixing term and the source term are changed;
in addition,  the  temperature  dependent  contributions, contained in the $\epsilon$-terms,   are  also gauge dependent.  Accordingly,
it is  convenient to  decompose the variation of the effective action into two parts and write  $\delta S_{eff } = \delta S_{0} + \delta S_{T}$. 
$\delta S_{0}$  arises  from  the variation of   the gauge fixing and source terms, and using (\ref{lef}), (\ref{sexpf}) and (\ref{sexpg})  it can be  worked out  as

\begin{equation}\label{vs0}
\delta S_{0} \left[ \bar{\Psi}_a ,  \Psi_a, A^\mu_a \right] 
=  \int dx  \left[ - \xi \epsilon_a  \left(\partial \cdot A_a \right)
 {\,\lower 1.0pt\vbox{\hrule \hbox{\vrule height 0.3 cm \hskip 0.28 cm \vrule height 0.3 cm}\hrule}\,}
   + J_{a}^{\mu} \partial_{\mu}
 - ie \left(\bar{\eta}_a \psi_a -\bar{\psi}_a \eta_a \right) 
\right] \Lambda_a 
\, , \end{equation}
with $\epsilon_1 =1 \, , \epsilon_2 =- 1$.
The quantity  $\delta S_{T}$ is obtained from the $\epsilon$-terms   in  (\ref{sexpf}) and (\ref{sexpg}). The result is 

\begin{eqnarray}\label{vst}
 && \delta S_{T}   \left[ \bar{\Psi}_a , \Psi_a, A^\mu_a \right] 
 =  i  \epsilon \int d \vec{x} dx_0 dy_0  \bigg \{ -
 \Lambda_a(x_0)  {\partial \over \partial x^\mu}{} \left[N_{ab}(x_0 - y_0) A^\mu_b(y_0) \right]
\nonumber \\ 
 & &+ i e \left[ \Lambda_a(x_0) -   \Lambda_b(y_0) \right] 
\bar{\psi}_a(x_0,\vec x) M_{ab}(x_0 - y_0) \psi_b(y_0,  \vec x)  \bigg \} 
 \, .   \end{eqnarray}

Although $S_{eff}$ varies  with the gauge transformations,
the physical consequences
of the  theory, expressed in terms of   Green's functions, should not  depend on the gauge. Thus,  the generating functional   $Z$ must be  gauge invariant and this 
non-trivial requirement leads to the Ward identities.
Combining  the previous results   with Eq. (\ref{zq2}) the gauge invariance of $Z$  implies 

\begin{equation}\label{giz}
\left \{\delta  S_{0} \left[{\delta  \over \delta  \eta_a }, 
{\delta  \over \delta \bar{\eta}_a }, {\delta  \over \delta J^\mu_a } 
 \right]  + \delta  S_{T} \left[{\delta  \over \delta  \eta_a }, 
{\delta  \over \delta \bar{\eta}_a }, {\delta  \over \delta J^\mu_a } 
 \right]   \right\} 
Z \left[J_\mu, \eta,\bar{\eta}\right]  \, = \, 0  \, . 
\end{equation}
By means of Eqs. (\ref{gam}), (\ref{rel1})  and using the results in 
(\ref{vs0}) and (\ref{vst})  we convert (\ref{giz}) into a  condition for 
 $\Gamma$:

\begin{eqnarray}\label{gigam}
&&  \!\!\!\!\!\!\!\!\!\!\!\!\!\!\!\!\!\!\!\!\!\!\!\! 
 -  \xi  \epsilon_a 
{\,\lower 1.0pt\vbox{\hrule \hbox{\vrule height 0.3 cm \hskip 0.28 cm \vrule height 0.3 cm}\hrule}\,}
 \partial \cdot A^a(x)   + 
 \partial^\mu { \delta \Gamma \over \delta A_a^\mu (x)}  -ie  \left( \psi_a(x)
{\delta \Gamma \over \delta \psi_a(x)}  -
 \bar{\psi}_a(x){\delta \Gamma \over \delta \bar{\psi}_a(x)} \right)  
\nonumber \\
+ i \epsilon e \int dy_0&&  \!\!\!\!\!\!
 \left[ i \bar{\psi}_a(x_0) M_{ab}(x_0 - y_0) \psi_b(y_0)
  + M_{ab}(x_0 - y_0){\partial  \psi_b(y_0) \over \partial \eta_a(y_0)} 
- (a \leftrightarrow b, x_0\leftrightarrow y_0 )\right]  
\nonumber \\
&-&  i \epsilon  \int dy_0  {\partial \over \partial x^\mu}{} \left[ N_{ab}(x_0 - y_0) A^\mu_b(y_0,\vec x)\right] 
 \, = \, 0  \, ,\end{eqnarray}
where we used   that the gauge functions $\Lambda_a(x)$  are
arbitrary, and no summation over the index $a$ is implied.
  
Equation  (\ref{gigam}) expresses  the  general content of the Ward identities
in the real-time formalism. 
Repeated differentiation of it, at 
$A^\mu_a = \bar{\psi}_a = \psi_a = 0$,  generates relations between 
the  one particle irreducible  Green functions that are  consequence of the gauge invariance of the 
theory. First we  differentiate (\ref{gigam})  with respect to $A_\mu(x)$    and using  the fact that 
$\delta \Gamma/ \delta A_\mu \delta A_\nu$ is the inverse of  the 
full  photon  propagator ${\bf D}_{\mu\nu}$, the following 
result  (in  momentum space) is obtained
 
\begin{equation}\label{wig1}
q^\mu{{\bf D}_{\mu\nu}}^{-1}(q) = \, - \, 
\xi  \sigma_3  \, q_\nu q^2     
 \,  - \, i \epsilon q_\nu \left[{\bf U}^{-1}(q_0) \right]^2 \, .
\end{equation}
This  relation for the full inverse photon  propagator holds to all orders in perturbation theory.
The effect of radiative 
corrections  is to add the  negative of the  self-energy   ${\bf \Pi}_{\mu\nu}$ to the inverse of the bare propagator $[{\bf D}^0_{\mu\nu}]^{-1}$. 
 From  the expression in  (\ref{invg}) it is easy to check that  
 (\ref{wig1}) is  satisfied by $[{\bf D}^0_{\mu\nu}]^{-1}$.
Consequently   we get

\begin{equation}\label{wig2}
q^\mu{{\bf \Pi}_{\mu\nu}} (q) \, = \, 0 \, . 
\end{equation}
In a recent work, the transversality of the polarization tensor at finite temperature has been  corroborated by an explicitly one-loop calculation \cite{das}. 
As we proved  here,    this  property   remains valid  to all orders in perturbation theory.

As a second application of the identity (\ref{gigam}), we  act on it  with 
$\delta^2/\delta\bar{\psi}(x) \delta \psi(y)$ and put  
$A^\mu_a = \bar{\psi}_a = \psi_a = 0$.  Using the  definitions of 
the proper vertex function (\ref{3pv}) and the full inverse   fermion propagator
(\ref{2pv}), we arrive to

\begin{equation}\label{wif1}
q_\mu \Gamma^\mu_{abc}(q,p) =
\delta_{ab} \left[ S^{-1}_{ca}(p + q)  - i \epsilon
 \left[ V^{-1}(p_0 + q_0) \right]^2_{ca} \right]
- \delta_{ac} \left[ S^{-1}_{ab}(p )  - i \epsilon \left[ V^{-1}(p_0 ) \right]^2_{ab} \right]
\, . \end{equation}
These are the Ward identities
that,  in the RTF,  relate the components of the full electron  propagator  to the vertex functions. Equation  (\ref{wif1})  represents eight
relations, one for each  element of  $\Gamma^\mu_{abc}$. 
They include   explicit temperature contributions due to the presence 
of the $\epsilon$-terms. 
At zero temperature, the matrix  $V(p_0 )$ reduces to the identity, 
the $\epsilon$ dependent terms  cancel,  and   (\ref{wif1}) decompose in two independent 
relations, one for the type-1 fields and another  for the  type-2 fields.
  
If we substitute the  inverse bare propagator (\ref{invf}),
Eq. (\ref{wif1}) simplifies  to

\begin{equation}\label{wif2}
q_\mu \Gamma^\mu_{abc}(q,p) = q_\mu  \gamma^\mu \epsilon_a \delta_{ab} \delta_{ca}
\, , \end{equation}
which  are  trivially satisfied, because  the only vertices  that  do 
not vanish at the  tree
level are  $\Gamma^\mu_{111} = - \Gamma^\mu_{222} = \gamma^\mu$. 
 According to this result,  it is convenient to write

\begin{equation}\label{vert}
 \Gamma^\mu_{abc}(q,p) =  \gamma^\mu \epsilon_a \delta_{ab} \delta_{ca}
+ \Lambda^\mu_{abc}(q,p) 
\, , \end{equation}
where $\Lambda^\mu_{abc}(q,p)$ represents the radiative corrections to 
the vertex function.  In terms of   the fermion self-energy matrix 
$ {\bf \Sigma}  (p)$, the inverse of the complete  fermion propagator can  be written
as ${\bf  S} ^{-1}   (p) =  \left[ {\bf  S}^{(0)}  (p) \right]^{-1}  -  {\bf \Sigma}  (p)$.
This relation  combined with  Eqs. (\ref{wif1})-(\ref{vert}) implies that

\begin{equation}\label{wif3}
q_\mu \Lambda^\mu_{abc}(q,p) = - 
\left[  \delta_{ab} \Sigma_{ca}(p + q)  
- \delta_{ac}  \Sigma_{ab}(p )   \right] \, . 
\end{equation}
Explicitly, these relations  are 

\begin{eqnarray}\label{wif4}
q_\mu \Lambda^\mu_{111}(q,p) &=& - \left[   \Sigma_{11}(p + q)  
-  \Sigma_{11}(p )   \right]   
\, , \nonumber \\
q_\mu \Lambda^\mu_{112}(q,p) &=& -   \Sigma_{21}(p + q)  
 \, , \nonumber \\
q_\mu \Lambda^\mu_{121}(q,p) &=&   \Sigma_{12}(p )  
  \, , \nonumber \\
 q_\mu \Lambda^\mu_{122}(q,p)  &=& 0  
 \, , \nonumber \\
q_\mu \Lambda^\mu_{211}(q,p)  &=& 0  \, , \nonumber \\
 q_\mu \Lambda^\mu_{212}(q,p)  &=&   \Sigma_{21}(p) 
\, , \nonumber \\
 q_\mu \Lambda^\mu_{221}(q,p)  &=& -  \Sigma_{12}(p + q) 
\, , \nonumber \\
q_\mu \Lambda^\mu_{222}(q,p) &=& - \left[   \Sigma_{22}(p + q)  
-  \Sigma_{22}(p )   \right]  
\,  .  \end{eqnarray}

Several  important  results can be read from the above  identities. 
First,  the vertex functions with equal  thermal fermion indices   but a different   photon index ($i.e.$ $\Lambda_{122}^\mu$ and
 $\Lambda_{211}^\mu$)  are  transverse.
Next,  in  the limit  $q_\mu \to 0$  the diagonal 
vertex functions
$\Lambda^\mu_{111}(0,p)$ and  $\Lambda^\mu_{222}(0,p)$  may be determined as follows in terms of   the diagonal components of the self-energy: 
 
\begin{eqnarray}\label{wif5}
\Lambda^\mu_{111}(0,p) &=&  - { \partial \Sigma_{11}(p) \over \partial p_\mu}
  \, , \nonumber \\
\Lambda^\mu_{222}(0,p) &=&  - { \partial \Sigma_{22}(p) \over \partial p_\mu} \, . 
\end{eqnarray}
Furthermore,  comparing the second (third) line  with the 
sixth (seven) line in  (\ref{wif4}) we see that the relations

\begin{eqnarray}\label{wirel}
\Lambda^\mu_{112}(q,p) \, = \, -  \, \Lambda^\mu_{212}(q,p + q) \, , 
\nonumber \\ 
\Lambda^\mu_{221}(q,p) \, = \, - \,  \Lambda^\mu_{121}(q,p+q) \, ,
\end{eqnarray}
are satisfied by the longitudinal components of the corresponding 
elements of $\Lambda^\mu_{abc}$. 
Notice that  the fermion   momentum in the right side is shifted from 
$p$ to $p + q$. The   four vertex functions in the last equations are imaginary since $\Sigma_{12}$ and $\Sigma_{21}$  in  Eqs (\ref{wif4}) are imaginary
quantities.

Equation 
(\ref{wif4})  implies that,   either the non-diagonal elements of $\bf \Sigma$ 
vanish  or that the vertex functions with different  values for the fermion  thermal index have a   singularity at $q^\mu = 0$.  Since  we know  that the self-energy is in general non-diagonal, we  conclude  that  
the four vertex function in (\ref{wirel})
 are   singular in the  $q_\mu \to 0$  limit. 
Later on,  this fact will be illustrated within a  one loop-calculation.

The foregoing relations can be  applied to simplify  the  longitudinal part of the   effective casual  vertex defined in (\ref{3pvr}).   First,  if  we combine the circling  equation (\ref{rel3}) with    (\ref{rel2})  and   (\ref{wirel}) we can derive   the following  relation 

\begin{equation}\label{nurel1}
  \, Im \,  \Lambda^\mu_{111}(q,p) =   i e^{-\beta \mu/2} \left[ \csc(2 \varphi_{p+q}) \Lambda^\mu_{121}(q,p+q)
-  \csc(2 \varphi_p) \Lambda^\mu_{121}(q,p) \right] \, ,
\end{equation}
for the longitudinal components.  Using these results we then find that the retarded  vertex functions  in  Eq. (\ref{3pvr}) can be written as

\begin{eqnarray}\label{3pvret}
 Re \Lambda^\mu_{R1}(q,p) &=& Re \Lambda^\mu_{R2}(q,p) = Re \Lambda^\mu_{R3}(q,p) = Re  \Lambda^\mu_{111}(q,p) \, ,  \nonumber \\
Im  \Lambda^\mu_{R1}(q,p) &=& {1 \over i} e^{-\beta \mu/2}  \left[ \csc(2 \varphi_{p+q}) \Lambda^\mu_{121}(q,p+q)
+ \csc(2 \varphi_p)  \Lambda^\mu_{121}(q,p)  \right] \, ,  \nonumber \\
Im  \Lambda^\mu_{R2}(q,p) &=& - Im  \Lambda^\mu_{R3}(q,p) =  {1 \over i}
e^{- \beta \mu/2} \left[ \csc(2 \varphi_{p+q})  \Lambda^\mu_{121}(q,p+q) -   \csc(2 \varphi_p)  \Lambda^\mu_{121}(q,p) \right] \, .  \nonumber \\ 
\end{eqnarray}
These equations are very useful because they  show that we need only  know 
one (complex)   of   eight vertex functions $\Lambda_{abc}$    to determine  the retarded vertex function of QED. In fact,  according to  (\ref{3pvret})  we can choose  to calculate 
 $Re \, \Lambda^\mu_{111}(q,p)$  and $Im \, \Lambda^\mu_{121}(q,p) $ to determine the other components of  $\Lambda^\mu_{abc}$ as well as the
three retarded vertex functions.
Although,  $\Lambda^\mu_{121}(q,p) $  is singular for $q_\mu  \to 0$, we notice 
that this singularity is  cancelled  for  
$  \Lambda^\mu_{R2}(q,p)$ and $  \Lambda^\mu_{R3}(q,p)$. Consequently,  
only the imaginary part of  $\Lambda^\mu_{R1}(q,p)$ presents a singularity at $q_\mu \to 0$.

Referring back to  Eqs. (\ref{sigret})  we see that the real and imaginary parts of  $\Sigma_R$ are obtained  from  $Re \Sigma_{11}$ and $\Sigma_{12}$ respectively. 
Here, we have derived  an analogous result for the retarded vertex functions.
Indeed,  in  (\ref{3pvret}) the real and imaginary  parts of the retarded 
vertex functions are expressed in terms of  $Re \Lambda_{111}$ and $\Lambda_{121}$ respectively.
However, it should be stressed  that whereas the result for   the fermion self-energy holds  in any theory,  Eq. (\ref{3pvret}) ensues  from  the gauge invariance of the theory.

Finally, combining (\ref{3pvr})   with  (\ref{wif4})
the Ward identities for the retarded vertex functions read 

\begin{eqnarray}\label{wif6}
q_\mu \Lambda^\mu_{R1}(q,p)  &=& - \left[\Sigma_A(p+q) - \Sigma_R(p)   \right]
\, , \nonumber \\ 
q_\mu \Lambda^\mu_{R3}(q,p)  &=&  q_\mu \Lambda^{\mu*}_{R2}(q,p) = 
- \left[\Sigma_R(p+q) - \Sigma_R(p)   \right]
\, , \end{eqnarray}
where $\Sigma_A = \Sigma_R^*$ is the advanced self-energy. 
Thus,  we notice that   the vertex function  with   the largest 
time corresponding to the  out-going  fermion  is only related 
to the retarded  self-energy. On the contrary,  if  the photon vertex has the largest  time, then the  Ward identity  relates the vertex to a combination of the advanced and  retarded self-energies. 
These equations are consistent with the fact that only 
$ \Lambda^\mu_{R1}(q,p)$ is singular  at $q_\mu = 0$.

Now, we explicitly verify  that  Ward identities are satisfied 
to the one-loop order.  To this order, the contribution  to the vertex function, represented by diagram $(a)$  in   Fig.  ~\ref{gura},  is 
 
\begin{equation}\label{vert1}
- i e \Lambda_{abc}^\mu  (q,p) = - (i e)^3 \epsilon_a \epsilon_b \epsilon_c  
\int {d^4 k \over (2 \pi)^4} {1 \over i} \left[D^{(0)}\right]^{\rho\sigma}_{bc}(k) \gamma_\sigma 
i S^{(0)}_{ac}(p+q - k) \gamma^\mu i S^{(0)}_{ba}(p - k) \gamma_\rho \, .
\end{equation}
The  fermion  and photon  propagators  are  the bare propagators   in Eqs.  (\ref{invf}) and (\ref{invg}), respectively. 

Similarly,  the first  contribution to the self-energy is shown in diagram $(b)$  of Fig.   ~\ref{gura} and reads

\begin{equation}\label{self1}
- i  \Sigma_{ab}  (q,p) = - (i e)^2   \epsilon_a \epsilon_b  
\int {d^4 k \over (2 \pi)^4} {1 \over i} \left[D^{(0)}\right]^{\rho\sigma}_{ba}(k) \gamma_\sigma 
i S^{(0)}_{ba}( {p} {\mkern -9.0mu /}  - k)  \gamma_\rho \, . 
\end{equation}

\let\picnaturalsize=N
\def\picsize{3.0in}
\def\picfilename{figu.eps}
\ifx\nopictures Y\else{\ifx\epsfloaded Y\else\input epsf \fi
\let\epsfloaded=Y
\centerline{\ifx\picnaturalsize N\epsfxsize \picsize\fi \epsfbox{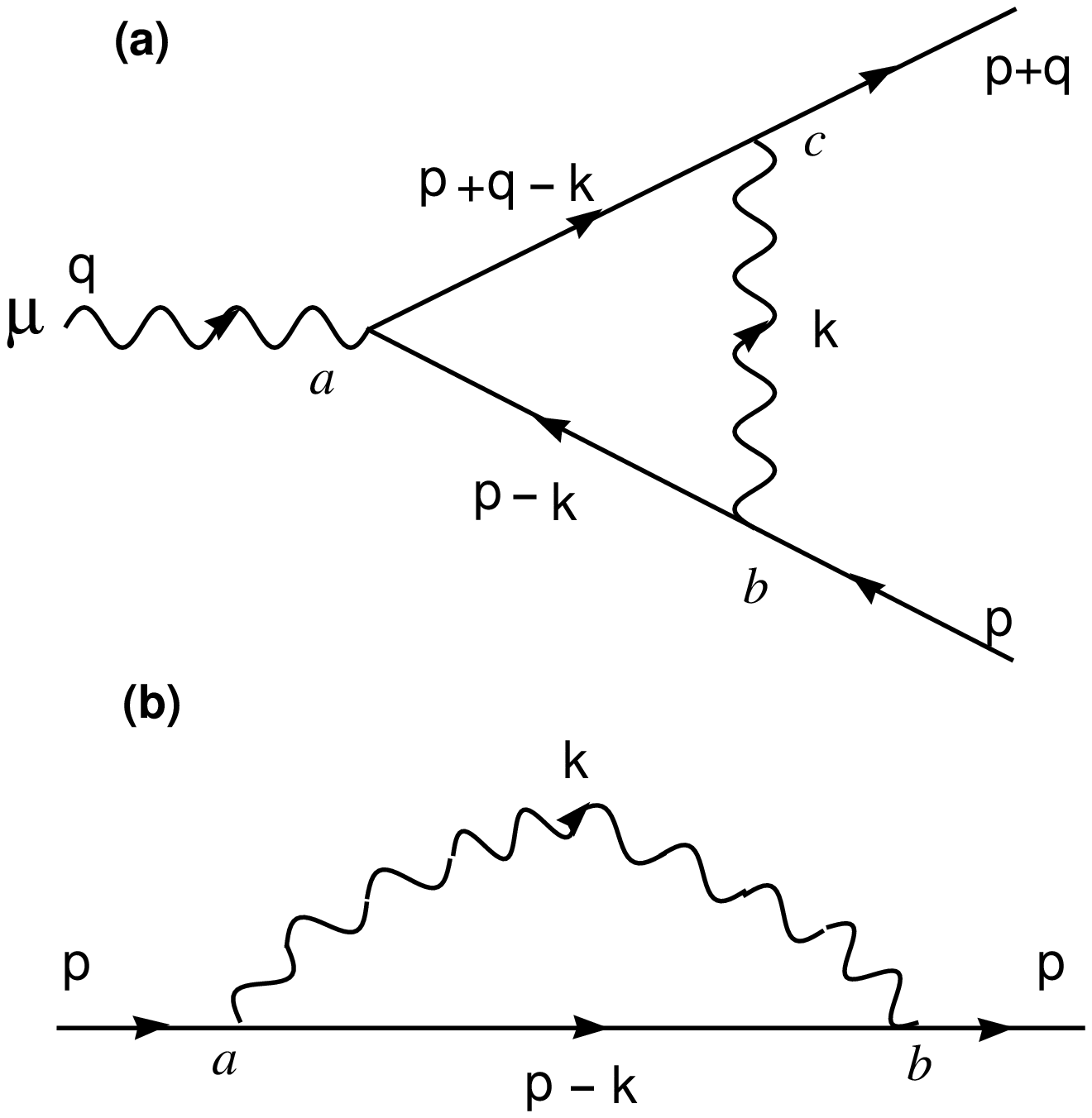}}}\fi

(a) The vertex $\Lambda^\mu_{abc}(q,p)$  to the one-loop order.
 (b) Fermion self-energy to the one-loop order.

\bigskip\bigskip

The  bare fermion propagator 
in  (\ref{invf})  satisfies the relation 

\begin{equation}\label{rel12}
 S_{ab}^{(0)} (p) \, \left( {p}  {\mkern -9.0mu /}  - m  \right) \, = \, \epsilon_a \delta_{ab}  +
 {\cal O}\left\{  \epsilon^2,  { \epsilon \, (p^2 - m^2) \over (p^2 - m^2)^2 + \epsilon^2}    \right\}  \,\,   \approx \epsilon_a \delta_{ab} 
\, .   \end{equation}
Here  we can  safely take  the limit $\epsilon \to 0 $, obtaining   well defined results.
These are  all that  we need.  If we  contract the vertex 
in (\ref{vert1}) with  $q_\mu$ ,   making the replacement 
${q}{\mkern -9.0mu /} = \left( {q}{\mkern -9.0mu /} + {p}{\mkern -9.0mu /}- {k}{\mkern -9.0mu /} - m \right)- \left({p}{\mkern -9.0mu /} - {k}{\mkern -9.0mu /} -m \right)$ and using  Eqs. (\ref{rel12}) and  (\ref{self1})  we immediately obtain the result given in Eq. (\ref{wif3}).
Hence,  the Ward identities in (\ref{wif3} and \ref{wif4}) are indeed 
verified  at the one loop-order. 

Finally, we  turn the  attention  to the origin of the  singular behavior of the vertex functions  $\Lambda_{abc}(q,p); \,  \, b\neq c$ in the  limit $q_\mu \to 0$.  Let us consider for example the function $\Lambda_{121}(q,p)$,  an easy algebraic
manipulation  of Eq. (\ref{vert1}) shows that the loop integral contains, among others, 
the   factors 
\begin{equation}\label{singu}
 \sin^2 \varphi_{p+q -k} \Delta^*( p+q - k) \Delta( p - k)  - 
 \cos^2 \varphi_{p+q -k} \Delta( p+q - k) \Delta^*( p - k) \, 
\end{equation}
 where $\Delta( p ) = 1/(p^2 - m^2 + i \epsilon) $ is the Feynman propagator.
Then, it is obvious that for any value  $q_\mu \neq 0$  
the integral is free of pinch singularities, or $\Delta^*(k) \Delta(k)$ terms.
However,  in the  $q_\mu  \to   0$ limit  the  integral develops a pinch  singularity. 
This is the origin, at the one-loop level, of the 
 singular behavior of the vertex function.    Nevertheless, it must be stressed
that we have proved that the singular behavior should remain valid to all 
orders in perturbation theory.
A detailed calculation of the one-loop vertex functions  $\Lambda_{abc}(q,p)$
with the corresponding   analysis  of the structure  of the singularity at
$q_\mu  =  0$  will be presented  elsewhere \cite{nos}.

\section{Conclusions}\label{conclu}

We have studied the  discrete and gauge symmetries  of  Quantum Electrodynamics at  finite temperature  within the  the real-time formalism.  
In the path integral representation of the generating functional the thermal information is inserted through boundary conditions, which are taken into account in  the $\epsilon$ factor. The presence of this convergence factor leads to an action that is non-local    in time, besides being   gauge dependent.
By demanding  the  generating functional to be  gauge independent we obtain  the  Ward identities,  which    relate the eight vertex functions to the elements  of the self-energy matrix. 
Combining the relations  obtained from the $Z_2$  symmetry  and the gauge symmetry 
 we find  that  (for  the longitudinal part)  only one out of   eight vertex functions is independent. In addition,   the retarded  vertex functions obey relations similar to those known for the  retarded self-energy. 

As a consequence  of the  Ward identities,  the vertex functions $\Lambda^\mu_{abc} $
with $b\neq c$  are found to be  singular when the photon momentum goes to zero.
 The zero-momentum limit of Feynman amplitudes at finite  temperature has  generated much discussion.  In particular,  it is known that  for  the thermal self-energy   calculated in perturbation theory,  the limits $p_0 \to 0$ and $|\vec p| \to 0 $ are not interchangeable. 
 It has been argued that the nonanalyticity of the self-energy  at $p_\mu = 0$  should be expected on physical grounds.
In particular Weldon \cite{weldon2}    presented various  methods to prove that
the self-energy  for scalar bosons  has a branch cut in $p_0$ 
and consequently the $p_0 \to 0$ and $|\vec p| \to 0 $ limits  do not commute.
On the other  hand, various attempts  have been made to  improve the usual perturbation expansion   in such a way that the self energy becomes analytic  at $p_\mu =  0$ \cite{unos}. 
Although our results refer to the 3-point function of QED, 
they may be   relevant  in the  understanding of   the zero-momentum limit 
of Feynman amplitudes at finite  temperature.  
The singularity of the  QED vertex function  at zero momentum 
occurs   because of the Ward identities, and  therefore is not a consequence
of the approximation made in a perturbative calculation. 
However,  it should be stressed  that in this paper  the singularity in the QED vertex functions
is   predicted  based on the Ward identities, whereas  this  kind of singular behavior is known to exist even for theories without gauge invariance.

\acknowledgments

This  work  was  supported  in part  by  the Universidad Nacional Aut\'onoma de M\'exico under  Grants  DGAPA-IN100694 and DGAPA- IN103895


\begin{thebibliography}{99}
\bibitem{gen}For reviews, see for  example: 
		N. P. Landsman and Ch. G. van Weert, Phys. Rep. {\bf 145}, 141   (1987 );
              T. Altherr, Int. J. Mod. Phys.  A{\bf  8}, 5605 (1993). 
\bibitem{matsu} T. Matsubara, Prog. Theor. Phys. {\bf 14}, 351 (1955).
\bibitem{ume} H.  Umezawa, H. Matsumoto and M. Tachiki, Thermo Field Dynamics
      and  Condensed  States (North-Holland, Amsterdam, 1982);  I. Ojima, 
     Ann. Phys. {\bf  137}, 1  (1981 );  H.  Umezawa, Advanced  Field Theory  
    (American   Institute of        Physics , New York, 1993); J. F. Nieves,
      Phys. Rev. D {\bf 42}, 4123 (1990).
\bibitem{niemi} A.J. Niemi and G.W. Semenoff, Nucl. Phys.  B {\bf 230}, 181 (1984); 
        Ann. Phys. (N.Y.) {\bf  152}, 105 (1984).
\bibitem{evans} T. S. Evans, Phys. Lett. B {\bf  252}, 108 (1990). 
\bibitem{kobes3} R. Kobes Phys. Rev. D {\bf 42}, 562 (1990); 
         Phys. Rev. D {\bf 43}, 1269 (1991). 
\bibitem{kobes} R. Kobes Phys. Rev. Lett. {\bf 67}, 1384 (1991).
\bibitem{koku}  R. Kobes, G. Kunstatter and A. Rebhan, Nucl. Phys.  B {\bf  355}, 1 
   (1991 );  R. Baier, G. Kunstatter and D. Schiff, Nucl. Phys.  B {\bf  388}, 
          287(1992).  
\bibitem{carri2} M. E. Carrington,  Phys. Rev. D {\bf  48}, 3836 (1993). In this paper,
self-consistent vertices for scalar QED in the  real time formalism  have  been constructed in order to satisfy the  Ward identities for  soft external momentum. 
\bibitem{braten}   E. Braaten and R. Pisarski, Nucl. Phys.  B {\bf  337}, 569 (1990 );
     Phys. Rev. Lett. {\bf  64}, 1338 (1990); J. Frenkel  and J. C. Taylor Nucl. Phys. 
      B {\bf 334},     199 (1990). 
\bibitem{kose}  A detailed derivation  of the  Feynman rules for  gauge fields and fermions  within the real-time  formalism  appears in: R. Kobes, G. W. Semenoff and N. Weiss, 
    Z. Phys. C {\bf  29}, 371 (1985).
\bibitem{das} A. Das and M. Hott, Mod. Phys. Lett. A {\bf  9}, 3383 (1994).
\bibitem{nos} J. D'Olivo,  M. Torres and E. T\'ututi, to be published. 
\bibitem{weldon2} H. A. Weldon,  Phys. Rev. D {\bf 47}, 594 (1993).
\bibitem{unos}  P. S. Gribosky and B. R. Holstein,  Z. Phys. C {\bf 47}, 205 (1990);
        J. Nieves and P. Pal,  Phys. Rev. D {\bf 51}, 5300 (1995).
\end{thebibliography}
\end{document}